\documentclass[12pt]{article}
\usepackage{epsfig}
\newcommand{\gev}{\mbox{GeV}}
\newcommand{\mev}{\mbox{MeV}}
        \newdimen\eqskip
        \newdimen\txtskip
        \baselineskip=\txtskip
        \newdimen\mysep                
        \newdimen\hmysep
\begin{document}
       
  \newcommand{\ccaption}[2]{
    \begin{center}
    \parbox{0.85\textwidth}{
      \caption[#1]{\small{{#2}}}
      }
    \end{center}
    }
\newcommand{\BS}{\bigskip}
\def    \be             {\begin{equation}}
\def    \ee             {\end{equation}}
\def    \beq             {\begin{equation}}
\def    \eeq             {\end{equation}}
\def    \ba             {\begin{eqnarray}}
\def    \ea             {\end{eqnarray}}
\def    \beqn           {\begin{eqnarray}}
\def    \eeqn           {\end{eqnarray}}
\def    \beeq           {\begin{eqnarray}}
\def    \eeeq           {\end{eqnarray}}
\def    \nn             {\nonumber}
\def    \=              {\;=\;}
\def    \frac           #1#2{{#1 \over #2}}
\def    \ret            {\\[\eqskip]}
\def    \ie             {{\em i.e.\/} }
\def    \eg             {{\em e.g.\/} }
\def \lsim{\mathrel{\vcenter
     {\hbox{$<$}\nointerlineskip\hbox{$\sim$}}}}
\def \gtrsim{\mathrel{\vcenter
     {\hbox{$>$}\nointerlineskip\hbox{$\sim$}}}}
\def    \bentarrow      {\:\raisebox{1.1ex}{\rlap{$\vert$}}\!\rightarrow}
\def    \rd             {{\mathrm d}}    
\def    \Im             {{\mathrm{Im}}}  
\def    \bra#1          {\mbox{$\langle #1 |$}}
\def    \ket#1          {\mbox{$| #1 \rangle$}}

\def    \kev            {\mbox{$\mathrm{keV}$}}
\def    \mev            {\mbox{$\mathrm{MeV}$}}
\def    \gev            {\mbox{$\mathrm{GeV}$}}


\def    \mq             {\mbox{$m_Q$}}  
\def    \mt             {\mbox{$m_t$}}  
\def    \mb             {\mbox{$m_b$}}  
\def    \mqq            {\mbox{$m_{Q\bar Q}$}}
\def    \mqqsq          {\mbox{$m^2_{Q\bar Q}$}}
\def    \pt             {\mbox{$p_T$}}
\def    \et             {\mbox{$E_T$}}
\def    \xt             {\mbox{$x_T$}}
\def    \xtsq           {\mbox{$x_T^2$}}
\def    \ptsq           {\mbox{$p^2_T$}}
\def    \etsq           {\mbox{$E^2_T$}}
\def    \rjet           {\mbox{$R_{\mathrm jet}$}}
\newcommand     \MSB            {\ifmmode {\overline{\rm MS}} \else 
                                 $\overline{\rm MS}$  \fi}
\def    \muf            {\mbox{$\mu_{\rm F}$}}
\def    \mug            {\mbox{$\mu_\gamma$}}
\def    \mufsq          {\mbox{$\mu^2_{\rm F}$}}
\def    \mur            {{\mbox{$\mu_{\rm R}$}}}
\def    \mursq          {\mbox{$\mu^2_{\rm R}$}}
\def    \mul            {{\mu_\Lambda}}
\def    \mulsq          {\mbox{$\mu^2_\Lambda$}}

\def    \bzero          {\mbox{$b_0$}}
\def    \as             {\ifmmode \alpha_s \else $\alpha_s$ \fi}
\def    \asmz             {\ifmmode \alpha_s(M_Z) \else $\alpha_s(M_Z)$ \fi}
\def    \asb            {\mbox{$\alpha_s^{(b)}$}}
\def    \assq           {\mbox{$\alpha_s^2$}}
\def \oacube {\mbox{$ {\cal O}(\alpha_s^3)$}}
\def \oaemacube {\mbox{$ {\cal O}(\alpha\alpha_s^3)$}}
\def \oafour {\mbox{$ {\cal O} (\alpha_s^4)$}}
\def \oatwo {\mbox{$ {\cal O} (\alpha_s^2)$}}
\def \oaematwo {\mbox{$ {\cal O}(\alpha \alpha_s^2)$}}
\def \oaemas {\mbox{$ {\cal O}(\alpha \alpha_s)$}} 
\def \oas   {\mbox{$ {\cal O}(\alpha_s)$}}
\def\slash#1{{#1\!\!\!/}}
\def\rt1{\raisebox{-1ex}{\rlap{$\; \rho \to 1 \;\;$}}
\raisebox{.4ex}{$\;\; \;\;\simeq \;\;\;\;$}}
\def\ltap{\raisebox{-.5ex}{\rlap{$\,\sim\,$}} \raisebox{.5ex}{$\,<\,$}}
\def\gtap{\raisebox{-.5ex}{\rlap{$\,\sim\,$}} \raisebox{.5ex}{$\,>\,$}} 

\newcommand\LambdaQCD{\Lambda_{\scriptscriptstyle \rm QCD}}

\def\herwig{\small HERWIG}
\def\isajet{\small ISAJET}
\def\pythia{\small PYTHIA}
\def\grace{\small GRACE}
\def\vecbos{\small VECBOS}
\def\madgraph{\small MADGRAPH}
\def\comphep{\small CompHEP}
\def\ALPGEN{\small ALPGEN}
\def\met{$\rlap{\kern.2em/}E_T$}
\begin{titlepage}
\nopagebreak
{\begin{flushright}{
 \begin{minipage}{5cm}
	CERN-TH/2003-054 \\
	hep-ph/0303085
\end{minipage}}\end{flushright}}
\vfill
\begin{center}
{ \bf \sc \Large The $t\bar{t}$ cross-section at 1.8 and 1.96 TeV:\\[0.5cm] 
a study of the systematics due to\\[0.5cm]
 parton densities and scale dependence\footnote{This work was
        supported in part by the EU Fourth Framework Programme
        ``Training and Mobility of Researchers'', Network ``Quantum
        Chromodynamics and the Deep Structure of Elementary
        Particles'', contract FMRX--CT98--0194 (DG 12 -- MIHT).}}
\\[1cm] {M. Cacciari$^{(a)}$, S. Frixione$^{(b)}$
  , M.L. Mangano$^{(c)}$, P. Nason$^{(d)}$ , G. Ridolfi$^{(b)}$ }
 \\[0.5cm] 
{\small 
$^{(a)}$ Dipartimento di Fisica, Universit\`{a} di Parma, Italy, and
  INFN, Sez. di Milano, gruppo collegato di Parma.\\
$^{(b)}$ INFN, Sezione di Genova, Italy. \\
$^{(c)}$ CERN, Theoretical Physics Division, Switzerland\\
$^{(d)}$ INFN, Sezione di Milano, Italy.
}
\end{center}                                   
\vfill
\begin{abstract}
We update the theoretical predictions for the $t\bar{t}$ production
cross-section at the Tevatron, taking into account the most recent
determinations of systematic uncertainties in the extraction of the
proton parton densities.
\end{abstract}
\vskip 3cm
CERN-TH/2003-054\hfill \\
March 6, 2003 \hfill  
\vfill       
\end{titlepage}

\section{Introduction}
We present in this note an update of the predictions for the top quark
production cross-section at the Tevatron. These  predictions
 are based on two complementary ingredients:
\begin{enumerate}
\item the evaluation of the parton-level cross-sections, carried out in
perturbative QCD with the inclusion of the full next-to-leading-order
(NLO) matrix elements~\cite{Nason:1987xz}, possibly improved with the
resummation to all orders of perturbation theory
 of classes of large soft logarithms~\cite{Sterman:1986aj,Catani:ne}
\item the proton parton densities (PDFs), which are typically
extracted comparing existing data with NLO calculations available for
the relevant processes, and extrapolated to the relevant region of
$Q^2$ using the NLO evolution equations (more recently, accurate
estimates of the exact NNLO results have also become
available~\cite{vanNeerven:2001pe}, based on partial evaluations of
the three-loop splitting functions).
\end{enumerate}
The numbers we present here are based on the theoretical framework
introduced in~\cite{Catani:1996yz} and~\cite{Bonciani:1998vc}, where
the complete NLO calculation of the $t\bar{t}$ cross-section was
improved with the resummation of leading~\cite{Catani:1996yz} and
next-to-leading~\cite{Bonciani:1998vc} soft logarithms appearing at all
orders of perturbation theory. The introduction of
resummation turns out  to have only a mild  impact on the overall
rates (the effects at NLL are typically of the order of a few percent),
but improves the stability of the predictions with respect to changes
of the renormalization scales.  While no progress has occurred since
1998 in the calculation itself, significant development has taken
place in the determination of the PDFs. 
In addition to much improved data from HERA, from fixed-target
DIS experiments at FNAL, and to the implementation of Tevatron jet and
$W$ production data in the fits, progress has occurred in the assessment
of the true uncertainties associated with the global fits to these
data. This work, which recently received considerable attention
(Giele, Keller and
Kosower~\cite{Giele:2001mr}, CTEQ~\cite{Stump:2001gu,Pumplin:2002vw}, 
MRST~\cite{Martin:2002aw}, Botje~\cite{Botje:1999dj},
Alekhin~\cite{Alekhin:2000es}), has led to sets of PDF parameterizations
which should provide a meaningful estimate of the ``1-$\sigma$''
uncertainty deriving from PDFs to be associated to any calculations of
hard processes in hadronic collisions.

The introduction of these PDF sets ``with uncertainties'' relaxes the
much constrained predictions which used to be anchored to predefined
functional parametrizations, and it is natural to anticipate that the
range of predictions for a given hard cross-section will be increased.

\section{Outline of the uncertainty estimate}
We shortly outline here the details of our  calculation, before presenting the
numerical results. Unless explicitly
 denoted as $\sigma_{NLO}$, all of our results are obtained using the 
 NLL-improved formalism of ref.~\cite{Bonciani:1998vc}.
\subsection{Scale uncertainty}
The evaluation of the purely theoretical uncertainty is based on the
standard exploration of the cross-section dependence on the
renormalization ($\mu_R$) and factorization ($\mu_F$) scales used in
the perturbative calculation. In this work, we follow the standard
convention of considering the range $m_{top}/2 < \mu < 2 m_{top}$,
setting $\mu_R=\mu_F\equiv \mu$. A
justification for this choice can be found in~\cite{Bonciani:1998vc},
where it was shown that $\mu \sim m_{top}/2$ corresponds to a point of
minimal sensitivity, providing a maximum of the cross-section in the
range $0.1<\mu/m_{top}<10$.  In the range of mass consistent with the
current data, and for the two CM energy values of run I and run II
(1.8 and 1.96~TeV, respectively), the relative scale uncertainty at
NLO is of the order of $\pm 10$\%, independent to good approximation
of $\sqrt{S}$, $m_{top}$ and PDF sets.  In this region of parameters,
the maximum value is obtained for $\mu\sim m_{top}/2$, and the minimum
for $\mu=2m_{top}$.  The inclusion of NLL resummation corrections
reduces the uncertainty to the level of approximately $\pm
5$\%~\cite{Bonciani:1998vc}\footnote{This number, as well as all
numerical estimates presented in this document, correspond to the
choice $A=2$, where $A$ is the parameter introduced
in~\cite{Bonciani:1998vc} to parameterize the uncertainty about
subleading higher order terms.  In that paper, it was found that $A=2$ gives
a better estimate of the higher order uncertainties. $A=0$, for
example, would reduce the scale dependence to only $\pm 2.5$\%,
without changing significantly the central value of the resummed
cross-section}. This is the effect of very small NLL corrections to
the NLO result for small values of $\mu$, where the NLO rate is
largest, and bigger corrections for large $\mu$.

For completeness, we also considered the possibility of varying
independently the value of renormalization and factorization
scale. These were chosen in the range $0.5< \mu_R/\mu_F <2$, with
$0.5<\mu_{R,F}/m_{top}<2$. We verified (see later)
 that within this range the results
obtained using the choice $\mu_R=\mu_F$  are not altered
significantly, leading only to a small increase of the upper estimate.

\subsection{PDF uncertainty}
In the framework of~\cite{Stump:2001gu,Pumplin:2002vw,Martin:2002aw},
 PDFs with uncertainties come in sets of $n_{PDF}$ pairs, 
where $n_{PDF}$ is the number of parameters used in the fits. Each
pair corresponds  to the fit obtained by varying of
$\pm 1\sigma$ the value of the fit parameter eigenvalues, after
diagonalization of the correlation matrix. By construction, the
systematic uncertainty obtained for the observable ${\cal O}$ is given
by:
\be \label{eq:err}
\Delta {\cal O} = \frac{1}{2} \; \sqrt {\sum_{i=1,n_{PDF}} \, ({\cal
  O}_{i+} - {\cal O}_{i-} )^2  }
\ee
where ${\cal O}_{i\pm}$ is the value obtained using the PDF set
corresponding to the variation of the $i$th eigenvalue within its
 error range.
The central value of the prediciton is obtained using a reference PDF
set, typically labelled with $i=0$.  We explore in this work the sets
in the CTEQ6~\cite{Pumplin:2002vw} parameterizations ($n_{CTEQ}=20$,
corresponding to 40 sets, plus 1 reference set) and in the MRST
2002~\cite{Martin:2002aw} compilation ($n_{MRST}=15$, corresponding to
30 sets, plus 1 reference set). All sets in the CTEQ compilation have
\asmz=0.118, while those in the MRST one have \asmz=0.119.  The CTEQ
sets are labeled as follows: {\bf 6M} for the default set, and {\bf
101}-{\bf 140} for the 20 $\pm 1 \sigma$ variations.  The MRST sets
are labeled as {\bf 0} for the reference set, and {\bf 1}-{\bf 30} for
the 15 $\pm 1 \sigma$ variations. In both cases, CTEQ and MRST, we use
 the default values of {\em tolerances} chosen by the two groups to
 best represent the uncertainty. In particular, CTEQ selects $\Delta
 \chi^2=100$, while MRST selects  $\Delta
 \chi^2=50$.

In addition, we shall also consider three sets obtained by the MRST
group in 2001~\cite{Martin:2001es}, where the values of \as\ was
frozen to $\pm 1\sigma$ from the central world average.  We shall
label these sets as {\bf A01L} for the low-\as (\asmz=0.117)
fit~\cite{Martin:2001es}, {\bf A01H} for the high-\as (\asmz=0.121)
fit~\cite{Martin:2001es}, {\bf J01} for a fit based on Tevatron jet
data~(\asmz=0.121)\cite{Martin:2001es}.

\section{Results}
Table~\ref{tab:mrsr}
summarizes the results obtained with the 
 PDF sets used in 1998, when the work in ref.~\cite{Bonciani:1998vc}
 appeared. The numbers agree with
what appears in Table~1 of that document.

{\renewcommand{\arraystretch}{1.2}
\begin{table}
\begin{center}
\begin{tabular}{l|ll|ll|ll} \hline
$\sqrt{S}$ &  $\mu=m_{top}/2$ & &  $\mu=m_{top}$ & &  $\mu=2m_{top}$ \\
(GeV)      & $\sigma_{NLO}$ & $\sigma_{res}$  & $\sigma_{NLO}$ &
     $\sigma_{res}$  & $\sigma_{NLO}$ & $\sigma_{res}$ \\ \hline
1800  & 5.17 & 5.19 & 4.87 & 5.06 & 4.32 & 4.69 \\
1960 & 6.69 & 6.71 & 6.31 & 6.56 & 5.61 & 6.11 \\
\hline 
\end{tabular}                                                                 
\ccaption{}{\label{tab:mrsr}  Cross-section
  predictions  (in pb) for the 1998 MRSR2 PDF and $m_{top}=175$~GeV.}
\end{center}                                         
\end{table} }

{\renewcommand{\arraystretch}{1.2}
\begin{table}
\begin{center}
\begin{tabular}{lll|ll} \hline
$\sqrt{S}$ & $m_{top}$ & $r_{\mu}$  &
 $\sigma_{ref}$({\bf 6M}) &
 $\Delta \sigma$
\\ \hline
  1800 &    170 &  0.5 &    6.22 &    0.42 \\  
  1800 &    170 &  1 &    6.10 &    0.40 \\  
  1800 &    170 &  2 &    5.66 &    0.37 \\  
  1800 &    175 &  0.5 &    5.29 &    0.35 \\  
  1800 &    175 &  1 &    5.19 &    0.33 \\  
  1800 &    175 &  2 &    4.81 &    0.31 \\  
  1800 &    180 &  0.5 &    4.52 &    0.29 \\  
  1800 &    180 &  1 &    4.43 &    0.28 \\  
  1800 &    180 &  2 &    4.11 &    0.26 \\  
  1960 &    170 &  0.5 &    7.97 &    0.57 \\  
  1960 &    170 &  1 &    7.83 &    0.54 \\  
  1960 &    170 &  2 &    7.29 &    0.49 \\  
  1960 &    175 &  0.5 &    6.82 &    0.47 \\  
  1960 &    175 &  1 &    6.70 &    0.45 \\  
  1960 &    175 &  2 &    6.23 &    0.42 \\  
  1960 &    180 &  0.5 &    5.86 &    0.40 \\  
  1960 &    180 &  1 &    5.75 &    0.38 \\  
  1960 &    180 &  2 &    5.35 &    0.35 \\  
\hline
\end{tabular}                                                                 
\ccaption{}{\label{tab:cteq} Range of cross-section
  predictions (in pb) for the CTEQ6 family of PDFs at a fixed scale
  $r_{\mu}=\mu/m_{top}$. $\sigma_{ref}$ refers to the central value,
  using the {\bf 6M} set, and $\Delta\sigma$ is the error, as defined in
  eq.~(\ref{eq:err}).}  \end{center}
\end{table} }

{\renewcommand{\arraystretch}{1.2}
\begin{table}
\begin{center}
\begin{tabular}{lll|ll} \hline
$\sqrt{S}$ & $m_{top}$ & $r_{\mu}$  &
 $\sigma_{ref}$({\bf 0}) &
 $\Delta \sigma$
\\ \hline
  1800 &    170 &  0.5 &    6.25 &    0.19 \\  
  1800 &    170 &  1 &    6.14 &    0.18 \\  
  1800 &    170 &  2 &    5.69 &    0.17 \\  
  1800 &    175 &  0.5 &    5.32 &    0.16 \\  
  1800 &    175 &  1 &    5.22 &    0.15 \\  
  1800 &    175 &  2 &    4.84 &    0.14 \\  
  1800 &    180 &  0.5 &    4.54 &    0.13 \\  
  1800 &    180 &  1 &    4.45 &    0.12 \\  
  1800 &    180 &  2 &    4.12 &    0.11 \\  
  1960 &    170 &  0.5 &    8.05 &    0.27 \\  
  1960 &    170 &  1 &    7.91 &    0.26 \\  
  1960 &    170 &  2 &    7.35 &    0.24 \\  
  1960 &    175 &  0.5 &    6.88 &    0.22 \\  
  1960 &    175 &  1 &    6.76 &    0.21 \\  
  1960 &    175 &  2 &    6.28 &    0.19 \\  
  1960 &    180 &  0.5 &    5.89 &    0.19 \\  
  1960 &    180 &  1 &    5.79 &    0.18 \\  
  1960 &    180 &  2 &    5.38 &    0.16 \\  
\hline
\end{tabular}                                                                 
\ccaption{}{\label{tab:mrst} Range of cross-section
  predictions (in pb) for the MRST family of PDFs at a fixed scale
  $r_{\mu}=\mu/m_{top}$. $\sigma_{ref}$ refers to the central value,
  using the {\bf 0} set, and $\Delta\sigma$ is the error, as defined in
  eq.~(\ref{eq:err}).}  \end{center}
\end{table} }

Table~\ref{tab:cteq} gives the central value and error for the CTEQ
 sets, for three values of the top mass (170, 175 and 180~GeV) and the two
 CM energies of interest ($\sqrt{S}=1800$ and 1960 GeV). We list
 the results obtained at the three reference values of the
 mass scale $r_{\mu}=\mu/m_{top}=0.5, \, 1, \, 2$.
Table~\ref{tab:mrst} provides the same information for the MRST
 sets.

\begin{figure}
\begin{center}
\epsfig{file=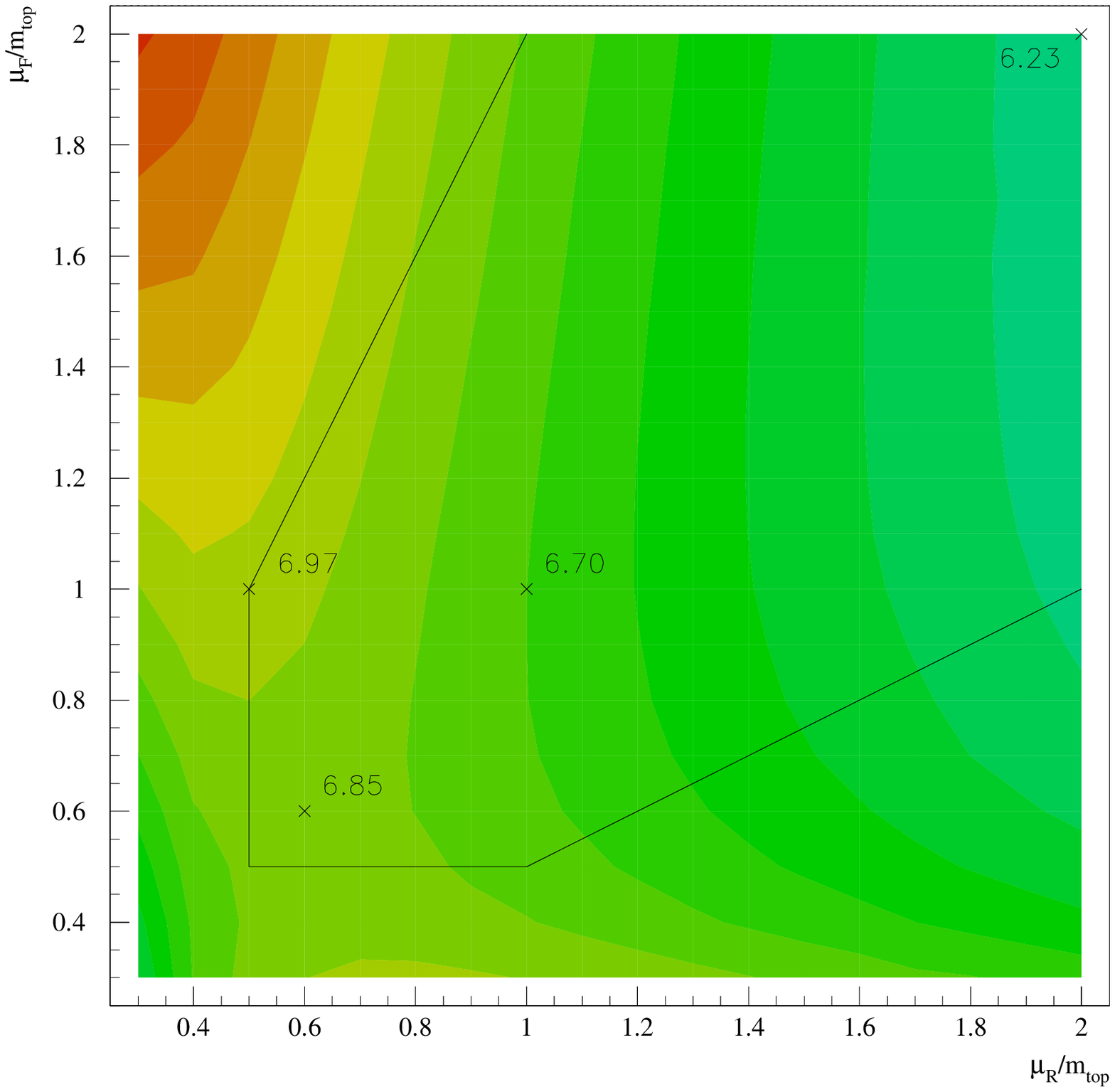,width=0.7\textwidth}
\vskip -2mm
\ccaption{*}{Contour plot of the NLL cross-section, in the
  $\mu_F-\mu_R$ plane. The oblique
solid line defines the region $0.5<\mu_R/\mu_F<2$.
\label{fig:muvar}
}
\end{center}
\end{figure}

Figure~\ref{fig:muvar} shows the contour plots of the NLL
cross-section when $\mu_R$ and $\mu_F$ are varied independently. The
region defined by the oblique solid lines corresponds to
$0.5<\mu_R/\mu_F<2$. It shows that within this domain the range of NLL
rates is compatible with the range obtained using $\mu_R=\mu_F$. 
 
In principle one should combine the uncertainty due to PDFs and that
 due to the scale choice in quadrature. We prefer to add them
 linearly, since the scale uncertainty is not really a systematic
 error in the strict sense. We therefore quote our range for the top
 cross-section as
\be  \label{eq:range}
\sigma(r_\mu=2)-\Delta\sigma_{PDF}(r_\mu=2) < \sigma < 
\sigma(r_\mu=1/2)+\Delta\sigma_{PDF}(r_\mu=1/2)
\ee
The corresponding values are given in Table~\ref{tab:cteqfull}.
The similar results for the MRST compilation are provided
 in~Table~\ref{tab:mrstfull}. 
{\renewcommand{\arraystretch}{1.2}
\begin{table}
\begin{center}
\begin{tabular}{ll|ccc} \hline
$\sqrt{S}$ & $m_{top}$ &
 $\sigma_{min}$ & $\sigma_{ref}$({\bf 6M}) & $\sigma_{max}$
\\ \hline
  1800 &    170 &    5.29 &    6.10 &    6.63 \\
  1800 &    175 &    4.51 &    5.19 &    5.64 \\
  1800 &    180 &    3.85 &    4.43 &    4.81 \\
  1960 &    170 &    6.79 &    7.83 &    8.54 \\
  1960 &    175 &    5.82 &    6.70 &    7.30 \\
  1960 &    180 &    5.00 &    5.75 &    6.25 \\
\hline
\end{tabular}                                                                 
\ccaption{}{\label{tab:cteqfull} Full range of cross-section
  predictions  (in pb) for the CTEQ6 family of PDFs, as defined in
  eq.~(\ref{eq:range}). $\sigma_{ref}$ refers to
  the choice of {\bf 6M} and $\mu=m_{top}$.}
\end{center}                                         
\end{table} }

{\renewcommand{\arraystretch}{1.2}
\begin{table}
\begin{center}
\begin{tabular}{ll|ccc} \hline
$\sqrt{S}$ & $m_{top}$ &
 $\sigma_{min}$ & $\sigma_{ref}$({\bf 0}) & $\sigma_{max}$
\\ \hline
  1800 &    170 &    5.52 &    6.13 &    6.44 \\
  1800 &    175 &    4.69 &    5.21 &    5.47 \\
  1800 &    180 &    4.00 &    4.44 &    4.67 \\
  1960 &    170 &    7.11 &    7.90 &    8.31 \\
  1960 &    175 &    6.08 &    6.76 &    7.10  \\
  1960 &    180 &    5.21 &    5.79 &    6.08 \\
\hline
\end{tabular}                                                                 
\ccaption{}{\label{tab:mrstfull} Full range of cross-section
  predictions  (in pb) for the MRST family of PDFs, as defined in
  eq.~(\ref{eq:range}). $\sigma_{ref}$ refers to
  the choice of set {\bf 0} and $\mu=m_{top}$.}
\end{center}                                         
\end{table} }

{\renewcommand{\arraystretch}{1.2}
\begin{table}
\begin{center}
\begin{tabular}{ll|ccc} \hline
$\sqrt{S}$ & $m_{top}$ & $\sigma_{min}$  ($r_{\mu}=2$, {\bf A01L}) &
 $\sigma_{ref}$($r_{\mu}=1$, {\bf 0})  &
 $\sigma_{max}$  ($r_{\mu}=0.5$, {\bf J01})
\\ \hline
1800 & 170 & 5.48 & 6.13 & 6.72 \\
1800 & 175 & 4.66 & 5.21 & 5.71 \\
1800 & 180 & 3.98 & 4.44 & 4.86 \\
1960 & 170 & 7.04 & 7.90 & 8.69 \\
1960 & 175 & 6.03 & 6.76 & 7.41 \\
1960 & 180 & 5.17 & 5.79 & 6.34 \\
\hline
\end{tabular}                                                                 
\ccaption{}{\label{tab:mrstas} Full range of cross-section
  predictions  (in pb) for the MRST family of PDFs. $\sigma_{ref}$ refers to
  the choice of {\bf 0} and $\mu=m_{top}$. $r_{\mu}=\mu/m_{top}$ and
  {\bf PDF}
  give the scale factor and PDF set  at which the minimum and maximum
  rates are attained.}
\end{center}                                         
\end{table} }

Three comments are in order:
\begin{enumerate}
\item the uncertainty ranges obtained using the CTEQ sets, for a fixed
  choice of scale, are almost twice as large as those for the MRST
  sets. We understand this is the result of the different tolerance
  criteria used by the two groups (see Appendix~B4
  of~\cite{Pumplin:2002vw} and Section~6 of~\cite{Martin:2002aw} for
  some discussion). 
 The MRST range increases however if we include in the analsys the 2001
  sets with varying \as. This is shown in Table~\ref{tab:mrstas}.  In
  this case the lowest predictions are obtained from the 2001 {\bf
  A01L} fit, with the low value of \as, while the highest prediction
  comes from the 2001 jet-based {\bf J01} fit. After the \as\
  variation is included, the MRST range becomes compatible with  that
  of CTEQ's.
\item the central values obtained today for the top cross-section are
 about 3\% larger than those obtained in 1998. At
$\sqrt{S}=1.8$~TeV and $\mu=m_{top}=175$~GeV we had 5.06~pb with the set
MRSR2 (\asmz=0.119). We now have 5.19~pb with CTEQ{\bf 6M},
 and 5.21~pb with MRST{\bf 0}. 
\item the contribution of the PDF systematics to the uncertainty range
  is large. In the case of the CTEQ sets, it is of the order of 6-7\%,
  larger than that due to the choice of scale. This is a
  result of the large sensitivity of the top cross-section to the
  large-$x$ gluon content of the proton, which is still poorly
  known. For CTEQ the largest contribution to the error comes from the
  two sets  {\bf 129} and {\bf 130}\footnote{This is consistent with
  what found in a recent study of jet produciton at the
  Tevatron~\cite{stump03}.}. 
 For these two sets, 
we find the contribution of the $gg$ channel to be
  respectively 11\% and 21\% of the total rate. For comparison, the
  contributions of the $q\bar{q}$ production channel for sets {\bf 129}
  and {\bf 130} are the same to within 1\%. In other words, the
  PDF uncertainty on the top rate is mostly driven by the poorly known
  gluon density, whose luminosity in this kinematic range varies by up
  to a factor of 2 within the 1-$\sigma$ PDF range.
\end{enumerate}
While the overall production rate has a large relative uncertainty of
approximately $\pm 15$\%, it is important to point out that the ratio
of cross-sections at $\sqrt{S}=1.96$~TeV and $\sqrt{S}=1.8$~TeV is
extremely stable. In the case of the CTEQ sets, for example, we found
$\sigma(1.96)/\sigma(1.8)=1.295\pm0.015$ after scanning
over the set of scale choices and for $170<m_{top}<180$~GeV.  The error is about 1\%.
We therefore consider the prediction
of the relative cross-section at the two energies to be a very stable
one.


For reference, we collect the full set of cross-sections (at
$\sqrt{S}=1.96$~TeV and $m_{top}=175$~GeV) for all CTEQ sets and scale
choices in Table~\ref{tab:cteqall}. Here, for the sake of
documentation, we provide the NLO rates and the NLL-improved ones
separately.

\section{Conclusions}
{\renewcommand{\arraystretch}{1.2}
\begin{table}
\begin{center}
\begin{tabular}{ll|ccc} \hline
$\sqrt{S}$ & $m_{top}$ &
 $\sigma_{min}$ & $\sigma_{ref}$({\bf 6M}) & $\sigma_{max}$
\\ \hline
  1800 &    170 &    5.29 &    6.10 &    6.72 \\
  1800 &    175 &    4.51 &    5.19 &    5.71 \\
  1800 &    180 &    3.85 &    4.43 &    4.86 \\
  1960 &    170 &    6.79 &    7.83 &    8.69 \\
  1960 &    175 &    5.82 &    6.70 &    7.41 \\
  1960 &    180 &    5.00 &    5.75 &    6.34 \\
\hline
\end{tabular}                                                                 
\ccaption{}{\label{tab:summary} Full range of cross-section
  predictions (in pb) for the combined study of CTEQ6, MRST and MRST with \as\
  variation. The central vlaues are taken from CTEQ6M. The minimum
  rates arise from CTEQ6, while the upper values arise from MRST set
  {\bf J01}. These numbers should be quoted as
  ``BCMN~\cite{Bonciani:1998vc}, as updated in [{\tt this paper}].''}
\end{center}                                         
\end{table} }
We reiterate here the main findings of this study. The inclusion of
the full PDF systematics, made possible by the recent works of several
groups, leads to a more realistic estimate of the top cross-section
uncertainty. The latest MRST and CTEQ sets give rise to cross-sections
which are typically 3\% larger than what obtained with sets available
at the time of Run~I. In addition to the increase in rate, the
size of the uncertainty range has also increased, to a value of the
order of $\pm 15$\%, dominated by the PDF and \as\ uncertainties.  The
leading source of PDF uncertainty comes from the (lack of) knowledge
of the gluon luminosity at large values of $x$. The $gg$ contribution
can in fact change through the PDF sets by up to a factor of 2 (from
10\% to 20\% of the total rate at 1.96~TeV). We find that the MRST
sets give rise to a smaller PDF uncertainty, a result we ascribe to
the tighter tolerances required by MRST in defining the range of the
eigenvalues. The MRST  uncertainty
increases however to values consistent with CTEQ's once the sets obtained from
a $\pm 1\sigma$ change of \asmz\ are included. This underscores the
importance of including the \as\ uncertainty into the PDF fits in a
more systematic fashion. On the same footing, the impact of higher
order corrections, as well as of the treatment of higher twist effects
in the fitting of low-$Q^2$ data, may need some more study before a
final tabulation of the PDF uncertainties is achieved~\cite{Martin:2002aw}.

We collect in Table~\ref{tab:summary} our final results. 
This summary table includes the CTEQ6M set and $\mu=m_{top}$ as
central values, and the most extreme rates extracted from
Tables~\ref{tab:cteqfull}, \ref{tab:mrstfull} and \ref{tab:mrstas} as
lower (with $\mu=2 m_{top}$) and upper values  (with $\mu=m_{top}/2$).

In spite of the overall large uncertainty, the ratio of
cross-sections at 1.96 and 1.8~TeV is extremely stable, being equal to
$1.295\pm 0.015$ over the mass range $170< m_{top} < 180$~GeV.

\section*{Acknowledgements} We thank S. Catani and J. Huston for
useful comments and discussions on the topic of this work. 

\pagebreak
{\renewcommand{\arraystretch}{1.0}
\begin{table}
\begin{center}
\begin{tabular}{l|ll|ll|ll} \hline
CTEQ6 & $\mu=m_{top}/2$ & &  $\mu=m_{top}$ & &  $\mu=2m_{top}$ \\
     & $\sigma_{NLO}$ & $\sigma_{res}$  & $\sigma_{NLO}$ &
     $\sigma_{res}$  & $\sigma_{NLO}$ & $\sigma_{res}$ \\ \hline
\hline
  {\bf 6M}  & 6.81 & 6.82 & 6.47 & 6.70 & 5.76 & 6.23\\
  {\bf 101} & 6.94 & 6.95 & 6.60 & 6.83 & 5.88 & 6.35\\
  {\bf 102} & 6.68 & 6.69 & 6.35 & 6.57 & 5.65 & 6.11\\
  {\bf 103} & 6.79 & 6.81 & 6.46 & 6.69 & 5.75 & 6.22\\
  {\bf 104} & 6.82 & 6.83 & 6.49 & 6.71 & 5.78 & 6.25\\
  {\bf 105} & 6.80 & 6.82 & 6.47 & 6.70 & 5.76 & 6.23\\
  {\bf 106} & 6.81 & 6.83 & 6.48 & 6.70 & 5.77 & 6.24\\
  {\bf 107} & 6.67 & 6.69 & 6.34 & 6.57 & 5.64 & 6.11\\
  {\bf 108} & 6.95 & 6.96 & 6.61 & 6.84 & 5.89 & 6.36\\
  {\bf 109} & 6.89 & 6.91 & 6.53 & 6.77 & 5.81 & 6.30\\
  {\bf 110} & 6.74 & 6.75 & 6.42 & 6.64 & 5.73 & 6.18\\
  {\bf 111} & 6.80 & 6.81 & 6.47 & 6.69 & 5.76 & 6.22\\
  {\bf 112} & 6.81 & 6.83 & 6.47 & 6.70 & 5.76 & 6.24\\
  {\bf 113} & 6.80 & 6.82 & 6.47 & 6.70 & 5.77 & 6.23\\
  {\bf 114} & 6.81 & 6.82 & 6.47 & 6.70 & 5.76 & 6.23\\
  {\bf 115} & 6.80 & 6.82 & 6.46 & 6.69 & 5.75 & 6.23\\
  {\bf 116} & 6.87 & 6.88 & 6.54 & 6.76 & 5.82 & 6.29\\
  {\bf 117} & 6.75 & 6.76 & 6.41 & 6.64 & 5.71 & 6.18\\
  {\bf 118} & 6.92 & 6.93 & 6.59 & 6.81 & 5.87 & 6.34\\
  {\bf 119} & 6.83 & 6.84 & 6.51 & 6.72 & 5.80 & 6.26\\
  {\bf 120} & 6.80 & 6.82 & 6.46 & 6.69 & 5.74 & 6.23\\
  {\bf 121} & 6.75 & 6.77 & 6.42 & 6.64 & 5.72 & 6.18\\
  {\bf 122} & 6.85 & 6.87 & 6.51 & 6.74 & 5.79 & 6.27\\
  {\bf 123} & 6.71 & 6.73 & 6.38 & 6.60 & 5.67 & 6.14\\
  {\bf 124} & 6.68 & 6.69 & 6.35 & 6.57 & 5.65 & 6.11\\
  {\bf 125} & 6.73 & 6.74 & 6.40 & 6.62 & 5.69 & 6.16\\
  {\bf 126} & 6.82 & 6.83 & 6.48 & 6.71 & 5.76 & 6.24\\
  {\bf 127} & 6.85 & 6.86 & 6.51 & 6.74 & 5.80 & 6.27\\
  {\bf 128} & 6.87 & 6.88 & 6.53 & 6.76 & 5.82 & 6.29\\
  {\bf 129} & 6.56 & 6.58 & 6.28 & 6.47 & 5.61 & 6.03\\
  {\bf 130} & 7.36 & 7.37 & 6.94 & 7.21 & 6.14 & 6.70\\
  {\bf 131} & 6.70 & 6.71 & 6.36 & 6.59 & 5.66 & 6.13\\
  {\bf 132} & 6.67 & 6.68 & 6.34 & 6.56 & 5.64 & 6.11\\
  {\bf 133} & 6.63 & 6.64 & 6.31 & 6.52 & 5.62 & 6.07\\
  {\bf 134} & 6.79 & 6.80 & 6.44 & 6.67 & 5.73 & 6.21\\
  {\bf 135} & 6.86 & 6.87 & 6.52 & 6.75 & 5.81 & 6.28\\
  {\bf 136} & 6.86 & 6.87 & 6.52 & 6.75 & 5.81 & 6.28\\
  {\bf 137} & 6.94 & 6.95 & 6.58 & 6.82 & 5.84 & 6.34\\
  {\bf 138} & 6.75 & 6.77 & 6.43 & 6.65 & 5.73 & 6.19\\
  {\bf 139} & 6.83 & 6.85 & 6.49 & 6.72 & 5.78 & 6.26\\
  {\bf 140} & 6.79 & 6.80 & 6.46 & 6.68 & 5.75 & 6.21\\
\hline
\end{tabular}                                                                 
\ccaption{}{\label{tab:cteqall} Full set of predictions for the CTEQ
  family of PDFs, and for $m_{top}=175$~GeV, at $\sqrt{S}=1.96$~TeV.
 $\sigma_{NLO}$ is the NLO rate, while $\sigma_{res}$ is the sum of
  NLO and NLL resummed, according to~\cite{Bonciani:1998vc}. All rates are in pb.}
\end{center}                                         
\end{table} }
\thispagestyle{empty}

\clearpage
\thispagestyle{plain}

\end{document}